\documentclass[12pt]{article}

\usepackage{amsmath,url}
\usepackage{amssymb}
\usepackage{epsfig}
\usepackage{color}
\usepackage{natbib}
  \usepackage{ccaption}
  \captionnamefont{\bfseries}
  \captiontitlefont{\raggedright\doublespacing}
  \captiondelim{. }
  
\usepackage[normalem]{ulem}
\usepackage{xkeyval} 
\usepackage{url}
\usepackage{setspace}

\usepackage{bm}
\bmdefine{\Bt}{t}
\bmdefine{\BX}{X}
\bmdefine{\BY}{Y}
\bmdefine{\BZ}{Z}
\bmdefine{\BB}{B}
\bmdefine{\BM}{M}
\bmdefine{\BD}{D}
\bmdefine{\Bi}{i}
\bmdefine{\Bj}{j}
\bmdefine{\Bx}{x}
\bmdefine{\By}{y}
\bmdefine{\Bz}{z}
\bmdefine{\Bw}{w}
\bmdefine{\Ba}{a}
\bmdefine{\Bb}{b}
\bmdefine{\Bc}{c}
\bmdefine{\Bh}{h}
\bmdefine{\Bg}{g}
\bmdefine{\Bu}{u}
\bmdefine{\Be}{e}

\newtheorem{thm}{Theorem}

\newtheorem{algorithm}[thm]{Algorithm}

\newtheorem{example}[thm]{Example}

\newcommand{\R}{\mathbb R}

\def\comment#1{\textit{[#1]}}
\def\comment#1{}

\definecolor{darkgreen}{rgb}{0,0.6,0}

\def\jdlqed{\vbox{\hrule \hbox{\vrule\hbox to
5pt{\vbox to 6pt{\vfil}\hfil}\vrule}\hrule}}

\begin{document}

\doublespacing

\noindent
BAYES ESTIMATORS
\vskip 0.2in

\doublespacing

\begin{center}
{\Large \bf Bayes estimators for phylogenetic reconstruction}
\vskip 0.2in

{P.M. Huggins$^1$, W. Li$^{2}$, D. Haws$^{3}$, T. Friedrich$^{3}$, J. Liu$^{2}$, and R. Yoshida$^{3}$\\
$^1${\it Lane Center for Computational Biology (Carnegie Mellon University)}\\
{\it Mellon Institute Building 4400 Fifth Avenue
Pittsburgh, PA 15213}\\
$^2${\it Department of Computer Science, The University of Kentucky, Lexington, KY, 40506-0046237}\\
$^3${\it Department of Statistics, University of Kentucky, Lexington, KY 40526-0027}\\
PMH, WL, and RY contributed equally to this work}
\end{center}

\noindent
Corresponding author: Ruriko Yoshida, \\
Department of Statistics, University of Kentucky, Lexington, KY 40526-0027\\
phone:(859) 257-5698, Fax:(859) 323-1973\\
email:\url{ruriko.yoshida@uky.edu},\\


\pagebreak

\raggedright
\hskip 0.5in {\it Abstract.--}
Tree reconstruction methods are often judged by their accuracy, measured by how close they get to the true tree. Yet most reconstruction methods like ML do not explicitly maximize this accuracy.  
To address this problem, we propose a Bayesian solution.  
Given tree samples, we propose finding the tree estimate which is closest on average to the samples. This ``median'' tree is known as the Bayes estimator (BE).  
The BE literally maximizes posterior expected accuracy, measured in terms of closeness (distance) to the true tree.  
We discuss a unified framework of BE trees, focusing especially on tree distances which are expressible as squared euclidean distances. 
Notable examples include Robinson--Foulds distance, quartet distance, and squared path difference.  
Using simulated data, we show Bayes estimators can be efficiently computed in practice by hill climbing.  
We also show that Bayes estimators achieve higher accuracy, compared to maximum likelihood and neighbor joining. 

\vskip 0.2in

\noindent
{\em key words}: Bayes estimator, consensus tree, path difference metric, phylogenetic inference.

\pagebreak

\vskip 0.8cm

\begin{center}
{\textsc{Introduction}}
\end{center}

\hskip 0.5in

When a large phylogeny is reconstructed from sequence data, 
it is typically expected that the reconstructed tree is at least slightly wrong, i.e. slightly different than the true tree.  
We refer to the difficulty in accurately reconstructing phylogenies as {\em tree uncertainty}.

\hskip 0.5in
Tree uncertainty is a pervasive issue in phylogenetics. 
To help cope with tree uncertainty, bootstrapping and Bayesian sampling methods provide a collection of possible trees instead of a single tree estimate.  
Using bootstrapping or Bayesian sampling, one common practice is to identify highly supported tree features (e.g. splits) which occur in almost all the tree samples.  
Highly supported features are regarded as likely features of the true tree.

\hskip 0.5in
Similarly, in simulation studies it is common to judge reconstruction methods based on how {\em close} they get to the true tree (\cite{Desper2003}).  Closeness to the true tree can be measured in many different ways.  One popular measure of closeness is the Robinson--Foulds (RF) distance (also known as symmetric difference).


\hskip 0.5in
These customary practices reflect a common view that when tree uncertainty is likely, a good reconstruction method ought to at least find a tree which is close to the true tree.  
For example, if multiple trees have high likelihood, then a good tree estimate should be an ``accurate representative'' of the high likelihood trees.  
Yet, reconstruction methods like maximum likelihood (ML) are not directly designed to achieve this goal.  This leads us to ask whether reconstruction accuracy (i.e. closeness to the true tree) can be improved, 
by attempting to directly optimize accuracy instead of likelihood.

\hskip 0.5in
Even though the true tree is unknown, we can still optimize reconstruction accuracy  using a Bayesian approach.  In the Bayesian view, the true tree is a random variable $T$ distributed according to the posterior distribution $P(T \, | \, D)$, where $D$ is input data such as sequence data.  If $d()$ measures distance between trees, and $T'$ is a tree estimate, then the expected distance between $T'$ and the true tree is ${\mathbb E}_{T \sim P(T \, | \, D)} d(T,T')$.  Thus, to maximize reconstruction accuracy, we should choose our tree estimate to be $T^* = \hbox{argmin}_{T'} {\mathbb E}_{T \sim P(T \, | \, D)} d(T,T')$ where $T^*$ is known as a {\em Bayes estimator}.   


\hskip 0.5in
Many popular distances between trees can be easily expressed as a squared euclidean distance, after embedding trees  in an appropriately chosen vector space.  
Important examples include Robinson--Foulds distance (symmetric difference), quartet distance, and the squared path difference.  In this paper, we focus on squared euclidean distances.

\hskip 0.5in
In statistical decision theory, Bayes estimators under squared euclidean distance are well understood and have nice properties.  
For example, under a squared euclidean distance, the Bayes estimator minimizes the distance to the mean of the posterior.  
This gives the result in \citep{Holder2008}:  The majority-rule consensus tree is the Bayes estimator, if closeness between trees is defined by Robinson-Foulds distance.   We also derive a closely related result for quartet distance:  Under quartet distance, the Bayes estimator tree is equivalent to a weighted quartet puzzling problem. 


\hskip 0.5in
In general, computing Bayes estimators is at least as hard as computing ML trees. Hill climbing techniques are popular and effective heuristics for hard tree optimization problems such as ML.  Thus we propose hill climbing to compute Bayes estimator trees as well.  For squared euclidean distances, each hill climbing step is quite fast, comparable to a traditional ML hill climbing step.  


\hskip 0.5in
We provide a simulation study of Bayes estimators using the path difference metric.  
We use hill climbing with nearest neighbor interchange (NNI) moves to find Bayes estimators.  We observe that hill climbing is fast in practice, after the preprocessing step of sampling the posterior on trees.  
More importantly, we observe that 
Bayes estimator trees are more accurate on average, compared to ML and neighbor joining (NJ).  These results comprise an encouraging pilot study of Bayes estimators.  
We conclude by discussing improvements and directions for future work developing Bayes estimators for phylogeny.

\begin{center}
{\textsc{Bayes estimators and squared euclidean distance}}
\end{center}

\hskip 0.5in
Let $D$ denote a collection of homologous sequences from $n$ species.  Many evolutionary models exist which express $P(D \, | \,T, \theta)$ in terms of an underlying phylogenetic tree $T$ on the n species, and evolutionary rate parameters $\theta$.  Given such a model, and observed sequence data $D$, there are two main methods for sampling trees $T$ which could have generated $D$:

\begin{itemize}
\item The Bayesian method, which declares a prior $P(T)$ on tree topologies, and uses sampling techniques such as Monte Carlo Markov Chain (MCMC) to approximately sample from $P(T \, | \,  D) \varpropto P(T) P(D \, | T)$,
\item The bootstrap method, which creates hypothetical datasets $D_i$ by bootstrapping columns from an alignment of $D$, and then computes a tree $T_i = T(D_i)$ for each $D_i$ by applying a tree reconstruction method such as ML or NJ.
\end{itemize}

\hskip 0.5in
The notation $P(T |  D)$ is not entirely appropriate for the distribution on trees obtained by the bootstrap method.  Nevertheless, for convenience we will use the notation $P(T  |  D)$ for the obtained distribution, regardless of whether the Bayesian or bootstrap method is used.

\hskip 0.5in
Given a measure of dissimilarity (or {\em distance}) $d(T, T')$ between phylogenetic trees on $n$ taxa, the {\em (posterior) expected loss} associated with a tree $T'$ is ${\mathbb E} d(T, T')$, where the expectation is taken over $T$, distributed as $P(T | D)$.  We write $\rho(T')$ for the expected loss.  The {\em Bayes estimator} $T^*$ minimizes the expected loss:
\[
 T^* = \hbox{argmin}_{T'} \, \rho(T')
\]
In other words, regarding the true tree $T$ as a random variable distributed as $P(T | D)$, the Bayes estimator is the tree $T^*$ which is closest to $T$ on average.

\hskip 0.5in
Bayes estimators are a common tool in statistical optimization and decision theory \citep{Berger1985}.  Given a finite sample $T_1, \ldots, T_N$ from $P(T | D)$, the {\em empirical expected loss} is $\hat{\rho}(T') = \frac{1}{N}  \sum_{i=1}^N d(T',T_i)$, and the empirical Bayes estimator is the tree that minimizes the empirical expected loss.  In this paper we will focus on empirical Bayes estimators for a given sample, and so we will simply say ``Bayes estimator'' when we mean the empirical Bayes estimator.

\vskip 0.8cm

\begin{center}
{{\it Squared euclidean distances between trees}}
\end{center}

\hskip 0.5in
Let $\mathcal{T}_n$ be the space of trees on $n$ taxa.  We call  $d(\cdot,\cdot)$ a {\em squared euclidean distance} if there is a function $v:\mathcal{T}_n \to {\mathbb R}^m$ for some $m$, such that
\[
d(T,T') = || v(T) - v(T') ||^2.
\]
We call $v()$ a (vector space) embedding.  Recall that for two vectors $a = (a_1, \ldots, a_m)$, $b = (b_1, \ldots, b_m)$ in ${\mathbb R}^m$, we have
$||a||^2 = \sum_{i=1}^m a_i^2$, and
$|| a - b ||^2 = ||a||^2 + ||b||^2 - 2(a \cdot b)$
where $a \cdot b$ denotes the dot product $\sum_{i=1}^m a_i b_i$.

\hskip 0.5in
Many popular distances between trees are squared euclidean distances.
Below we list several such distances, all of which were studied in \citep{Steel1993}. 
For each distance, we illustrate the vector space embedding and the distance using the two trees $T_1$ and $T_2$ shown in Figure \ref{fig3} (no branch lengths) and Figure \ref{fig4} (branch lengths).  



\begin{example}
Let $S(T)$ denote the set of splits induced by a tree $T$.
The (normalized) Robinson-Foulds distance \citep{Robinson1981} $d_{RF}(T,T')$ is half the size of the symmetric difference
$(S(T) - S(T')) \cup (S(T') - S(T))$.  The Robinson--Foulds distance can also be realized as the squared euclidean distance
\[
d_{RF}(T',T) = \frac{1}{2} || v_{RF}(T) - v_{RF}(T') ||^2
\]
where $v_{RF}: {\mathcal T}_n \to {\mathbb R}^{2^{n-1}-1}$ maps tree $T$ to the 0/1 vector $v_{RF}(T)$ whose nonzero entries correspond to splits in $T$.
For example, for the trees $T_1$ and $T_2$ in Figure \ref{fig3}, we have 
$$ v_{RF}(T_1) = (1,1,1,1,1,1,0,0,0,0,0,1,0,0,0), $$
$$ v_{RF}(T_2) = (1,1,1,1,1,1,0,0,1,0,0,0,0,0,0), $$
and 
$$ d_{RF}(T_1,T_2) = \frac{1}{2} || v_{RF}(T_1) - v_{RF}(T_2) ||^2 = 1.$$
Here the coordinates of $v_{RF}(T_1)$ and $v_{RF}(T_2)$ are given by
\begin{multline*} \Big(\{A\},\{B\},\{C\},\{D\},\{E\},\{A,B\},\{B,C\},\{A,C\},\{C,D\},\\
        \{B,D\},\{A,D\},\{D,E\},\{C,E\},\{B,E\},\{A,E\}\Big)
\end{multline*}
where for example $\{B,D\}$ corresponds to the partition $\{\,\{B,D\},\{A,C,E\}\,\}$.
\end{example}

\begin{example}
Let $Q(T)$ denote the set of quartets induced by a tree $T$.   The quartet distance \citep{Estabrook1985}
$d_Q(T,T')$ is half the size of the symmetric difference $(Q(T) - Q(T')) \cup (Q(T') - Q(T))$.
Analogous to the Robinson--Foulds distance, $d_Q$ can be realized as a squared euclidean distance,
\[
d_{Q}(T',T) = \frac{1}{2} || v_{Q}(T) - v_{Q}(T') ||^2
\]
where $v_{Q}: {\mathcal T}_n \to {\mathbb R}^{3 {n \choose 4}}$ maps tree $T$ to the 0/1 vector $v_{Q}(T)$ whose nonzero entries correspond to quartets in $T$.
For example, for the trees $T_1$ and $T_2$ in Figure \ref{fig3}, we have 
$$ v_{Q}(T_1) = (1,0,0,1,0,0,1,0,0,1,0,0,1,0,0), $$
$$ v_{Q}(T_2) = (1,0,0,0,0,1,1,0,0,0,0,1,1,0,0), $$
and
$$ d_{Q}(T_1,T_2) = \frac{1}{2} || v_{Q}(T_1) - v_{Q}(T_2) ||^2 = 2.$$
Here the coordinates of $v_{Q}(T_1)$ and $v_{Q}(T_2)$ are given by following cherry groupings (two leaves with the same parent node)
{\small
\begin{multline*} \Big(\, \{AB,CD\}, \{AC,BD\}, \{AD,BC\}, \{BC,DE\}, \{BD,CE\}, \{BE,CD\},\{AB,CE\}, \{AC,BE\}, \\
        \{AE,BC\}, \{AC,DE\}, \{AD,CE\}, \{AE,CD\}, \{AB,DE\}, \{AD,BE\}, \{AE,BD\}\, \Big).
\end{multline*}
}
\end{example}

\begin{example}
For $T \in {\mathcal T}_n$, let $D_T \in {\mathbb R}^{n \choose 2}$ be the matrix of pairwise distances between leaves in $T$.  The {\em squared dissimilarity map distance} is defined as $d_{D}(T',T) = ||D_T - D_{T'}||^2$.  The dissimilarity map distance is perhaps one of the oldest studied, see e.g. \citep{Buneman1971}.
For example, for the trees $T_1$ and $T_2$ in Figure \ref{fig4}, we have 
$$ D_{T_1} = (5.3,9.0,15.2,12.4,6.1,12.3,9.5,10.8,8.0,8.0), $$
$$ D_{T_2} = (3.5,11.3,13.2,10.9,12.0,13.9,11.6,7.1,7.0,8.9), $$
and
$$ d_{D}(T_1,T_2) = || D_{T_1} - D_{T_2} ||^2 = 72.06.$$
Here the coordinates of $D_{T_1}$ and $D_{T_2}$ are given by
$$ \Big( D_{1,2}, D_{1,3,}, D_{1,4}, D_{2,3}, D_{2,4}, \ldots, D_{4,5} \Big), $$
where $D_{i,j}$ is the sum of branch lengths of the path from leaf $i$ to $j$.
\end{example}

\begin{example}\label{ex2}
The distances $d_{RF}(T,T')$ and $d_Q(T,T')$ are {\em topological}  distances, i.e. they only depend on the topologies of $T,T'$, and not edge lengths.  The dissimilarity map distance does depend on edge lengths, but it has a natural topological analog called the {\em path difference metric.}  The squared path difference is
\[
d_{p}(T',T) = || v_{p}(T) - v_{p}(T') ||^2
\]
where $v_{p}(T) \in {\mathbb R}^{n \choose 2}$ is the integer vector whose $ij$th entry counts the number of edges between leaves $i$ and $j$ in $T$.  Path difference was studied in  \citep{Steel1993}.  Note that in our notation, we have squared the norm, whereas \citep{Steel1993} defined $d_{p}(T',T) = || v_{p}(T) - v_{p}(T') ||$. 
 
For example, for the trees $T_1$ and $T_2$ in Figure \ref{fig3}, we have 
$$ v_{p}(T_1) = (2,3,4,4,3,4,4,3,3,2), $$
$$ v_{p}(T_2) = (2,4,4,3,4,4,3,2,3,3), $$
and
$$ d_{p}(T_1,T_2) = || v_{p}(T_1) - v_{p}(T_2) ||^2 = 6.$$
Here the coordinates of $v_{p}(T_1)$ and $v_{p}(T_2)$ are given by
$$ \Big( v_{1,2}, v_{1,3,}, v_{1,4}, v_{2,3}, v_{2,4}, \ldots, v_{4,5} \Big), $$
where $v_{i,j}$ is the number of edges between leaf $i$ and $j$.
\end{example}



\hskip 0.5in
The above examples highlight the fact that many combinatorial distances can be interpreted as squared euclidean distances. Under a squared euclidean distance, the Bayes estimator is the projection of the mean onto the nearest tree.
More specifically, if $d(T,T') = || v(T) - v(T') ||^2$  is a squared euclidean distance, then evidently
\[
{\rho}(T') = ||v(T') - {\mu}||^2 + Var
\]
where ${\mu} = {\mathbb E} [v(T)]$ and ${\mu}_2 = {\mathbb E}[ \, ||v(T)||^2 \, ]$,
and $Var = {\mu}_2 - ||{\mu}||^2$ does not depend on $T'$.

\hskip 0.5in
For example, under the Robinson--Foulds distance, the Bayes estimator is obtained by projecting the vector of split frequencies $\mu_{RF} = {\mathbb E} v_{RF}(T)$ onto the nearest 0/1 vector  $v_{RF}(T^*) \in \{ v_{RF}(T')\}_{T'}  \subset \{0,1\}^{2^{n-1}-1}$.   
If we relax this problem, and simply project $\mu_{RF}$ onto the nearest 0/1 vector $v^* \in \{0,1\}^{2^{n-1}-1}$, then we see $v^*$ is obtained by rounding all 
entries in $\mu_{RF}$ to the nearest integer 0 or 1.  In other words $v^* = v_{RF}(T^*)$ where $T^*$ is the consensus tree.  
Thus we have  the result in \cite{Holder2008}:  
the consensus tree is the Bayes estimator for Robinson-Foulds distance.







\hskip 0.5in
In our view, projecting a point (e.g. input dissimilarity map) to a nearby tree is a geometric analog of a Bayes estimator.  Indeed, distance-based tree reconstruction methods can be loosely regarded as ``projections''
of an input dissimilarity map $D \in {\mathbb R}^{n \choose 2}$ onto a tree metric $D_T = D - \epsilon$, where $\epsilon$ is ``small'' according to some norm.  The geometry of distance-based tree reconstruction methods has been studied before, see \citep{kord2009,yoshida2008,Mihaescu2007}.  
\vskip 0.8cm

\begin{center}
{\textsc{Relation between Bayes estimators and existing reconstruction methods}}
\end{center}

\begin{center}
{{\it Quartet puzzling}}
\end{center}

\hskip 0.5in
Under the quartet distance $d_Q(T,T') = ||v_Q(T) - v_Q(T')||^2$, the Bayes estimator is the tree $T^*$ which minimizes 
$|| v_Q(T) - \mu_Q ||^2$, where $\mu_Q = {\mathbb E} v_Q(T)$ is the vector of posterior quartet frequencies.  Since $||v_Q(T)||^2 = {n \choose 4}$ for all trees on $n$ taxa, we have 
\[
|| v_Q(T) - \mu_Q ||^2 = {n \choose 4} + ||\mu_Q||^2 - 2 v_Q(T) \cdot \mu_Q = (constant) - 2 v_Q(T) \cdot \mu_Q
\]
and so the Bayes estimator $T^*$ can be equivalently defined as  $T^* = \hbox{argmax}_T \, \mu_Q \cdot v_Q(T)$.  Maximizing $\mu_Q \cdot v_Q(T)$ is a {\em weighted quartet puzzling} problem:  Given a set of weights $\mu_Q$ on quartets, find a compatible set of quartets of maximal weight.  If all quartet weights are 0/1, then we obtain the traditional quartet puzzling problem \citep{Strimmer1996}. 

\hskip 0.5in
Analogous to split frequencies and the consensus tree, we can use a sample of trees to estimate quartet frequencies, and then apply weighted quartet puzzling to find the Bayes estimator tree.  In general though, quartet puzzling (and hence weighted quartet puzzling) is NP-hard \cite{Steel1992}.  However, there has been considerable progress toward solving large instances: see \cite{warnow,Snir2009} for example.  In our case, the weights $\mu_Q$ have special structure since they are realizable by a collection of trees; this might make the weighted quartet puzzling we are considering here easier.

\begin{center}
{{\it Ordinary Least Squares (OLS) minimum evolution (ME)}}
\end{center}

\hskip 0.5in
For a squared distance dissimilarity map, there is a striking similarity between Bayes estimators and the minimum evolution (ME) approach to phylogenetic reconstruction.  ME methods are distance-based methods that have been extensively studied \citep{Holder2003,Rzhetsky1993}.  One of the earliest examples is Ordinary Least Square (OLS)  ME \citep{EDWARDS1963,Desper02fastand}.  OLS ME first estimates the branch lengths for each tree topology $T$ by minimizing $||D_T - D||^2$, where $D$ is the input dissimilarity map.  Then the outputted tree topology $T^*$ is the topology whose sum of estimated branch lengths is minimal.  If $D = D_T + \epsilon,$ where $D_T$ is a tree metric and $\epsilon$ comprises $i.i.d.$ errors with mean $0$, then OLS ME is statistically consistent as a method to recover $D_T$.

\hskip 0.5in
There is however a key difference between OLS ME and minimizing the expected squared dissimilarity map distance.  The input to OLS ME is a dissimilarity map presumed to be of the form $D = D_{T} + \epsilon$.  In sharp contrast, the mean $\mu$ summarizes the posterior distribution on $D_T$, given input such as sequence data.  Although $\mu$ could be viewed as a random variable whose distribution is governed by the true underlying tree $T$, the form of this distribution $P(\mu \, | T)$ is opaque and depends on the model of sequence evolution being used.  Thus, while directly minimizing $||D_T - \mu ||^2$ produces the Bayes estimator $T^*$, it is not clear whether the minimum evolution approach (treating $\mu$ as a ``perturbed tree metric'') is a sensible alternative.



\vskip 0.8cm

\begin{center}
{\textsc{Hill climbing optimization}}
\end{center}

\hskip 0.5in
Since the number of tree topologies on $n$ taxa grows exponentially in $n$, computing the Bayes estimator $T^*$ under a general distance function can be computationally hard.  However, hill climbing techniques such as those used in ML methods \citep{phyml} often work quite well in practice for tree reconstruction.  Hill climbing techniques can similarly be used to find local minima of the empirical expected loss. 

\hskip 0.5in
 Hill climbing requires a way to move from one tree topology
to another. Three types of combinatorial tree moves are often used for this purpose; {\em Nearest Neighbor Interchange (NNI)}, {\em Subtree-Prune-and-Regraft (SPR)},
and {\em Tree-Bisection-Reconnect (TBR)} \citep{Semple2003}.
SPR and TBR moves are more general than NNI, but every SPR and TBR move is a composition of at most two NNI moves.  SPR and TBR moves endow each tree with $O(n^2)$ neighbors.  NNI moves produce a smaller set of $O(n)$ neighbors. See (\cite{Allen2001}) for details. 
{\tt PHYML} uses NNI moves when hill climbing to quickly search for a ML tree \citep{phyml}. We follow their example and choose NNI moves to apply hill climbing.  




\hskip 0.5in
For each proposed move $T^{current} \to T^{new}$ during hill climbing, $\hat{\rho}(T^{new})$ must be computed.  A straightforward evaluation using the definition  $\hat{\rho}(T^{new} )= $ $\frac{1}{N} \sum_{i=1}^N d(T^{new}, T_i)$ requires $N$ evaluations of $d()$, where $N$ is the sample size.   For squared euclidean distances the situation is often much better since $\hat{\rho}(T^{new})$ can be re-expressed (up to an additive constant), as simply $\hat{\rho}(T^{new}) = d(\hat{\mu}, T^{new})$, where $\hat{\mu} = \frac{1}{N} \sum_{i=1}^N v(T_i)$ is the sample mean.  Note $\hat{\mu}$ 
does not depend on the tree $T^{new}$, thus it can be computed once at the beginning of hill climbing. Consequently, at each step we need only evaluate $d()$ once.
The computational expense to calculate $d()$ depends on the choice of vector space embedding.

\begin{center}
{\textsc{Simulation study:  Methods}}
\end{center}




\hskip 0.5in
For studying Bayes estimators, a natural first choice for distance between trees is Robinson--Foulds distance.  But then the Bayes estimator is the consensus tree, which has been extensively studied, and is easy to compute from samples.  We thus sought out other important distances besides Robinson--Foulds.

\hskip 0.5in
The dissimilarity map distance is one of the oldest distances for the comparison of
trees, and lies at the foundation of distance-based reconstruction methods.  Thus, dissimilarity map and related distances are a natural choice for case-study of Bayes estimators.  We specifically chose the (squared) path difference metric.  The path difference metric
$||v_{p}(T) - v_{p}(T')||$ 
is precisely the dissimilarity map distance $||D_T - D_{T'}||$, if all edge lengths in $T,T'$ are redefined to be 1.  
Setting all edge lengths to 1 prevents deemphasis of the shorter (presumably uncertain) edges.  
Intuitively, this emphasizes topological accuracy in the Bayes estimator.  We believe this is a desirable property, and we are not the first to suggest its importance.
The conclusion of \cite{Steel1993} states
\begin{quote}
``The path difference metric, $d_p$, has several interesting features that suggest that it merits more study and consideration for use when studying evolutionary trees.  These features will make it particularly attractive when studying large trees. $\cdots$ The $d_p$ metric may be the method of choice when trees are more dissimilar than expected by chance.'' 
\end{quote}
\hskip 0.5in
Thus we chose the squared path difference as a case study.  
We think quartet distance would also be interesting, but believe that a study of Bayes estimators under quartet distance should include quartet puzzling methods, given the close connections outlined in the Quartet Puzzling section.  
We have therefore deferred study of quartet distance.

\hskip 0.5in
Under the path difference metric, trees are embedded in $\R^{n \choose 2}$.  Using depth-first search on a tree $T$, the embedding vector $v(T)$ can computed in $O(n^2)$ time.  
Euclidean distance in  $\R^{n \choose 2}$ can also be computed in $O(n^2)$ time.


\begin{center}
{{\it Simulated data}}
\end{center}

\hskip 0.5in
For simulated data, we used the first $1000$ examples from the data set presented in \cite{phyml}.
We briefly review the details of the data set.  Trees on $40$ taxa were generated according to a Markov process.  For each generated tree, 40 homologous sequences (no indels) of length $500$ were generated, under the Kumura two-parameter (K2P) model \citep{Kimura1980}, with a transition/transversion
ratio of $2.0$.  Specifically the Seq-Gen program \citep{seqgen} was used to generate the sequences.
The data is available from the website
\url{http://www.atgc-montpellier.fr/phyml/datasets.php}.

\begin{center}
\noindent{\it Reconstruction methods}
\end{center}

\hskip 0.5in
For each set of homologous sequences $D$ in the simulated data, we used the software {\tt MrBayes} \citep{Mrbayes} to obtain $15000$ samples from the posterior distribution $P(T | D)$.  Specifically, we ran {\tt MrBayes} under the K2P model, discarded the initial $25\%$ of samples as a burn-in, used a $50$ generation sample rate, and ran for $1,000,000$ generations in total. 

\hskip 0.5in
We computed a ML tree estimate for each data set, using the hill climbing software {\tt PHYML} \citep{phyml} as described in the paper.  We also computed a NJ tree using the software {\tt PHYLIP} \citep{Felsenstein1989}, using pairwise distances computed by {\tt PHYLIP}. 

\hskip 0.5in
We then used our in-house software to minimize the expected path difference squared euclidean distance by hill climbing.  We performed hill climbing using NNI moves, along with various choices of starting trees. For starting trees we used the NJ tree, the ML tree, and five samples from $P(T | D)$ (NJ and ML trees were computed as described above).  We also used the {\tt MrBayes} tree sample which had the highest likelihood, which we call the ``empirical MAP'' tree.

\hskip 0.5in
We now briefly describe our hill climbing implementation.  The input for the algorithm is a list of trees $T_1, \ldots, T_N$ sampled from $P(T \, | \, D)$, and an initial starting tree $T^0$.  The pseudo-code is as follows:

\begin{algorithm}[Hill climbing from an initial tree $T^0$]

\begin{itemize}
\item[]
\end{itemize}

\begin{itemize}

\item[] INPUT:  Samples $T_1, \ldots, T_N$, and an initial tree $T^0$.
\item[] OUTPUT:  Local minimum $T^*$ of the empirical expected loss.

\item[] PROCEDURE:

\begin{itemize}
\item[] BEGIN
\item[] Compute and store $\hat{\mu}_{p} = \sum_i v_{p}(T_i)$.
\item[] Initialize $T^{*} = T^0$, and $\rho_{p}^* = || v_{p}(T^0) - \hat{\mu}_p ||^2$.
\item[] DO:
\begin{itemize}
\item[] Pick an NNI neighbor $T^{new}$ of $T^{*}$.
\item[] Compute $\rho_{p}^{new} = || v_{p}(T^{new}) - \hat{\mu}_p ||^2$.
\item[] IF $\rho_{p}^{new} < \rho_{p}^*$:
\begin{itemize}
\item[] Set $T^* = T^{new}$ and $\rho_{p}^* = \rho_{p}^{new}$.
\end{itemize}
\item[] END IF
\end{itemize}
\item[] UNTIL  \, $\rho_{p}^* < \rho_{p}^{new}$ is satisfied for all neighbors $T^{new}$ of $T^*$.
\item[] Output $T^*$
\item[] END

\end{itemize}

\end{itemize}
\end{algorithm}

\hskip 0.5in
In practice, allowing the hill climbing algorithm to run until complete convergence might take too long.  Thus, we included several alternative stopping criteria in the $UNTIL$ statement. (For example, halt if a maximum number of loop iterations is reached.)   In our simulation study, the algorithm always found a local maximum before halting.
The source code, written in java, is available at \url{http://cophylogeny.net/research.php}.

\vskip 0.8cm

\begin{center}
{\textsc{Simulation study:  Results}}
\end{center}

\begin{center}
{{\it Comparing objective functions for tree reconstruction }}
\end{center}

\hskip 0.5in
In our distance-based framework, the canonical measure of reconstruction accuracy is the distance,  $d_p(T^*, T^{true}) = || v_{p}(T^*) - v_{p}(T^{true})||$, between the true tree $T^{true}$ and the estimated tree $T^*$.  When reconstructing a tree, ideally we would like to directly use distance to the true tree as the objective function.  But obviously this is impossible unless $T^{true}$ is known.  One obvious question is:  How good are other objective functions, such as likelihood and $\hat{\rho}_{p}$, as proxies for $d_{p}(\cdot , T^{true})$?  The relationships among objective functions are particularly important for nearly optimal trees. 

\hskip 0.5in
We explored this question using the simulated data.  For each of the $1,000$ data sets, we computed three scores for each of the $15,000$ {\tt MrBayes} samples $T_i$, $i = 1, \ldots, 15000$.  The three scores we investigated are  1) The observed frequency of the tree topology in {\tt MrBayes} samples, 2) The empirical expected loss:  $\hat{\rho}_{p}(T_i)$ $= ||v_{p}(T_i) - \frac{1}{15000} \sum_j v_{p}(T_j)||^2$, and 3) The actual distance to the true tree:  $d_{p}(T_i, T^{true})$ $= || v_{p}(T_i) - v_{p}(T^{true})||^2$. 

\hskip 0.5in
For each data set, we restricted our attention to the $25$ most frequent tree topologies. The number of samples $15,000$ was large enough so that the frequencies of the $25$ most probable tree topologies could be estimated fairly well in most cases.  For the $25$ most probable topologies, we computed the Kendall-tau correlations between the three scores and recorded the results in Table \ref{table3}.

\hskip 0.5in
If there are no ties among the $25$ topologies under any of the scores, then the Kendall-tau has a natural interpretation:   If $P(s_2(T) < s_2(T') | s_1(T) < s_1(T') = p$ for a randomly drawn pair $T,T'$ of the $25$ topologies, then the Kendall-tau correlation is $2p - 1$ between the scores $s_1,s_2$.  As Table \ref{table3} shows, our proposed empirical expected loss $\hat{\rho}_{p}$ outperforms likelihood, as a proxy for the distance to the true tree.


\begin{center}
{{\it Performance of tree reconstruction methods }}
\end{center}

\hskip 0.5in
As described in (Simulation Study:  Methods), for each simulated data set we computed NJ, ML, and empirical MAP trees.  We then performed NNI-based hill climbing to optimize $\hat{\rho}_{p}$, using NJ/ML/MAP as starting trees as well as starts chosen randomly from {\tt MrBayes} samples.  We estimated the Bayes estimator (BE) tree by taking the best of five random starts.


\hskip 0.5in
Following \cite{phyml}, we plotted the inaccuracy (path difference to true tree) of the NJ, ML, empirical MAP, and BE trees (Figure \ref{fig1}).  Notice we have reported the inaccuracy between trees $T,T'$ as the norm $||v_{p}(T) - v_{p}(T')||$, instead of the square $||v_{p}(T) - v_{p}(T')||^2$.  We chose to do this so that the inaccuracy can be loosely interpreted as ``average difference of number of edges between a typical pair of leaves.''  In the plot, inaccuracy is plotted against the maximum unadjusted pairwise
divergence in the sequence data. The unadjusted pairwise divergence between two sequences is the proportion of sites where both sequences
differ.

\hskip 0.5in
We also give an analogous plot (Figure \ref{fig2}), plotting the empirical expected loss $\hat{\rho}_{p}(T)$ for the various tree estimators.
Note the true tree might not be the global optimum of $\hat{\rho}_{p}(T)$.  Thus we included the true tree in the plot as well.

\hskip 0.5in
Tables (\ref{table1}) and (\ref{table2}) summarize the results of our NNI-based hill climbing when ML/NJ/empirical MAP trees are used as the starting tree.  Note the ML tree (computed by phyML) was obtained by NNI hill climbing optimizing the likelihood.  Our hill climbing optimizes $\hat{\rho}_{p}$ instead, so it is possible an NNI move can improve the phyML tree.

\hskip 0.5in
We indeed observed that NJ, ML, and empirical MAP trees can be improved by hill climbing.  (Table \ref{table2}) and (Table \ref{table1}) give summary information.  In particular, (Table \ref{table1}) shows that our hill climbing algorithm improves the distance to the true tree.  We find this particularly encouraging.

\hskip 0.5in
Using a Pentium dual core system running Red Hat Linux 4, each run of our hill climbing programs required between $1$ minute and $1.5$ minutes on average per example, depending on the starting tree.  Using the NJ tree as the initial tree took longer on average, because more hill climbing steps were required to find a local optimum.

\begin{center}
{\textsc{Discussion}}
\end{center}


\hskip 0.5in
For phylogenetic reconstruction, the Bayes estimator is a natural choice when recovering the true tree is unlikely, and one is content to find a tree which is ``close'' to the true tree.
Here ``close'' is defined by a choice of distance between trees, e.g. Robinson--Foulds distance.  The Bayes estimator directly maximizes its expected accuracy, measured in terms of closeness to the true tree.  
In contrast, ML optimizes likelihood instead of accuracy.

\hskip 0.5in
As observed in \cite{Holder2008}, the popular consensus tree has a natural interpretation as the Bayes estimator which minimizes the expected Robinson--Foulds distance to the true tree.
Thus, for the special case of Robinson--Foulds distance, Bayes estimators have actually been studied for quite some time.


\hskip 0.5in
As part of an exploratory simulation study, we showed that hill climbing can be used to find an empirical Bayes estimator in practice, given a sample of trees from the posterior distribution.  
In particular we used the squared {\em path difference metric} described in \cite{Steel1993}.
Hill climbing optimization produced tree estimates which were closer to the true tree, outperforming NJ and ML.  
And in the majority of cases, hill climbing improved distance to the true tree, even when the initial tree was obtained by hill climbing optimization of the likelihood.  
We consider this very encouraging for future work on hill climbing approaches for Bayes estimators.

\hskip 0.5in
Systematists are best qualified to help choose which types of distances should be used to compare trees.  
On the theoretical front, some interesting new distances are being studied  such as the geodesic distance \citep{citeulike:3063901,Owen2008,Owen2009,Owen2009b}.  
We believe Bayes estimators (or ``median trees'') under novel distances comprise an interesting direction for future mathematical research. 
We also think Bayes estimators under the classical quartet distance might be interesting, in light of the close connection to quartet puzzling.



\hskip 0.5in
In this paper we used NNI moves to apply the hill climbing algorithm.  One could also try more general tree moves such as SPR or TBR, analogous to \citep{Gascuel2005a}. 
It would be interesting to study which tree moves give faster hill climbing convergence for Bayes estimators in practice.  
Similarly, exploration strategies such as Tabu search \citep{Glover1986Future-Paths-fo} or simulated annealing may give better performance.

\hskip 0.5in

\hskip 0.5in
For some vector space embeddings (e.g. quartet embedding $v_Q()$), the embedding vectors for trees may be rather high-dimensional and non-sparse.  
Then it may be faster to use the naive definition $\hat{\rho}(T) = \frac{1}{N} \sum_{i=1}^N d(T,T_i)$ directly.  
Indeed, quartet distance $d_Q(T,T')$ can be computed in $O(n \log n)$ time for two trees on $n$ taxa \citep{Pedersen2001}, which is much faster than operations on the vectors $v_Q(T), v_Q(T')$ which have dimension $O(n^4)$.


\hskip 0.5in
In this paper, we have focused on different types of tree {\em features} that can be used to define distance, e.g. splits or quartets. 
Systematists are particularly interested in splits.  
Thus, one could also study different ways to define a distance based on splits. 
For example,  \cite{Holder2008} considered a generalized Robinson--Foulds distance that allows a specificity/sensitivity trade-off.
We think another interesting way to modify Robinson--Foulds distance would be to make the distance more ``local''.  
For example, one could define a transformed distance $d(T,T') = \min (d_{RF}(T,T'), K)$ for a given ``ceiling'' constant $K > 0$. 
Then the Bayes estimator could be interpreted as a ``smoothed'' ML tree, i.e. the ML tree after the likelihood has been smoothed by a local convolution. 
This smoothed ML tree could provide a nice compromise between ML trees and consensus trees.

\hskip 0.5in
Finally, we note that in our simulation study, ML trees were quite accurate.  In fact, the ML tree was typically quite close to the Bayes estimator, in terms of NNI moves.  
Thus an ML (or approximate ML) tree might be quite useful as an initial guess for a Bayes estimator tree.  
Then, one could ``polish'' the MLE by using hill-climbing optimization of the expected loss.

\begin{center}
\noindent{\textsc{Acknowledgments}}
\end{center}

\hskip 0.5in
The authors would like to thank D. Weisrock for the many useful comments which improved this paper.  The second, the third, the fourth, and last authors are supported by NIH Research Project Grant Program (R01) from the Joint DMS/BIO/NIGMS Math/Bio Program (1R01GM086888-01 and 5R01GM086888-02).  
%
%
%

\pagebreak
\bibliographystyle{sysbio}
\doublespacing

\bibliography{tree_bayes_est}

\pagebreak

\begin{table}[!ht]
\begin{center}
\caption{ 
Using {\tt MrBayes} under the K2P model for $1,000,000$ generations sampling every $50$ generations.
We then removed the first $5,000$ sampled trees out of 20,000 sampled trees as 25\% burn-in to obtain 15,000 samples from the posterior distribution $P(T | D)$. 
We give the performance of hill climbing, applied to several different initial trees. The first two columns summarize how the local minimum compared to the initial tree, on the 1000 simulated data sets.  The third column gives the average percentage by which hill climbing decreases the path difference distance to the true tree.  This is computed as $1 - $ mean($d_{initial} / d_{final}$), where mean() denotes denotes geometric mean.  If either the initial or final distance to the true tree is zero, we add 1 to both distances.  
} \label{table1}
{\small
\begin{tabular}{lccc} \hline
Initial tree & Hill climbing improves & Hill climbing worsens & Avg drop in \\
             & distance to $T^{true}$? & distance to $T^{true}$? & distance to $T^{true}$\\ 
ML tree & 380 & 253 & $5.9$\% \\
Empirical MAP tree  & 508 & 185 & $17.9$\% \\
NJ tree &  693 & 229 & $39.6$\% \\ 
\\ \hline
\end{tabular}
}
\end{center}
\end{table}

\pagebreak
\begin{table}[!ht]
\begin{center}
\caption{
Using {\tt MrBayes} under the K2P model for $1,000,000$ generations sampling every $50$ generations.
We then removed the first $5,000$ sampled trees out of 20,000 sampled trees as 25\% burn-in to obtain 15,000 samples from the posterior distribution $P(T | D)$. 
 The first two columns summarize how the local minimum compared to the initial tree, on the 1000 simulated data sets.  The third column gives the average percentage by which hill climbing decreases $\hat{\rho}_{p}$.  This is computed as $1 - $ mean($\sqrt{\hat{\rho}_{p}^{initial} / \hat{\rho}_{p}^{final}}$), where mean() denotes denotes geometric mean.  If either $\hat{\rho}_{p}^{initial}$ or $\hat{\rho}_{p}^{final}$ is zero, we add 1 to both.   
} \label{table2}
{\small
\begin{tabular}{lccc} \hline

Initial tree & Hill climbing improves & Hill climbing worsens & Avg drop in \\
             & $\hat{\rho}_{p}$? &  $\hat{\rho}_{p}$? &  $\hat{\rho}_{p}$ \\ 
ML tree &  690 & 0 & $5.9$\% \\
Empirical MAP tree & 870 & 0 & $8.6$\% \\ 
NJ tree & 961 & 0 & $20.3$\% \\ \\\hline
\end{tabular}
}
\end{center}
\end{table}

\pagebreak

\begin{table}[!ht]
\begin{center}
\caption{For each of the $1,000$ data sets, we computed three scores for each of the $15,000$ {\tt MrBayes} samples $T_i$, $i = 1, \ldots, 15000$.  The three scores we investigated are  1) The observed frequency of the tree topology in {\tt MrBayes} samples, 2) The empirical expected distance $\hat{\rho}_{p}(T_i) = ||v_{p}(T_i) - \frac{1}{15000} \sum_j v_{p}(T_j)||^2$, and 3) The actual distance $d_{p}(T_i, T^{true}) = || v_{p}(T_i) - v_{p}(T^{true})||^2$.}\label{table3}
\vskip 0.2in

\begin{tabular}{rrrr}
 & $P(T_i)$ & $\hat{\rho}_{p}(T_i)$ & $d_{p}(T_i, T^{true})$ \\ 
\hline
$P(T_i)$ & $\cdot \, $ & 0.352 & 0.148 \\
$\hat{\rho}_{p}(T_i)$ & & $\cdot$ \,  & 0.270 \\
$d_{p}(T_i, T^{true})$ & & & $\cdot$ \, \\
\end{tabular}
\end{center}
\end{table}



\pagebreak

\begin{center}
{{\it Legends to Figures}}
\end{center}

\begin{figure}[ht!]
\begin{center}
 \end{center}
\caption{Normalized, unsquared path difference ${||v_{p}(T^{true}) - v_{p}(T^*)||} \,\,\, / \, {\sqrt{n \choose 2}}$ for tree estimates $T^*$ computed by various reconstruction methods, for $1000$ simulated trees $T^{true}$ on $n = 40$ taxa.  Here NJ (N) is the neighbor joining tree constructed via {\tt neighbor} in {\tt PHYLIP} package, ML (L) is the {\tt PHYML} tree, MAP (M) is the {\tt MrBayes} sample with the highest posterior probability, and Bayes (B) is the Bayes Estimator tree, estimated from {\tt MrBayes} samples.  } \label{fig1}
\end{figure}

\begin{figure}[ht!]
\begin{center}
 \end{center}
\caption{Normalized empirical expected distance to the true tree, $\sqrt{\frac{1}{N} \sum_{i=1}^N ||v_{p}(T^{true}) - v_{p}(T^*)||^2} \,\,\, / \, \sqrt{n \choose 2}$ for tree estimates $T^*$ computed by various reconstruction methods, for 1000 simulated trees $T^{true}$ on $n = 40$ taxa.  Here NJ (N) is the neighbor joining tree constructed via {\tt neighbor} in {\tt PHYLIP} package, ML (L) is the {\tt PHYML} tree, MAP (M) is the {\tt MrBayes} sample with the highest posterior probability, and Bayes (B) is the Bayes Estimator tree, estimated from {\tt MrBayes} samples.  } \label{fig2}
\end{figure}

\begin{figure}[ht!]
\caption{Two trees on five taxa with different topologies, $T_1$ (left) and $T_2$ (right).} \label{fig3}
\end{figure}

\begin{figure}[ht!]
\caption{Same two trees in Figure \ref{fig3} with branch lengths assigned, $T_1$ (left) and $T_2$ (right).} \label{fig4}
\end{figure}

\pagebreak

{\bf Figure 1.}

\begin{figure}[ht!]
\includegraphics[scale= 0.6]{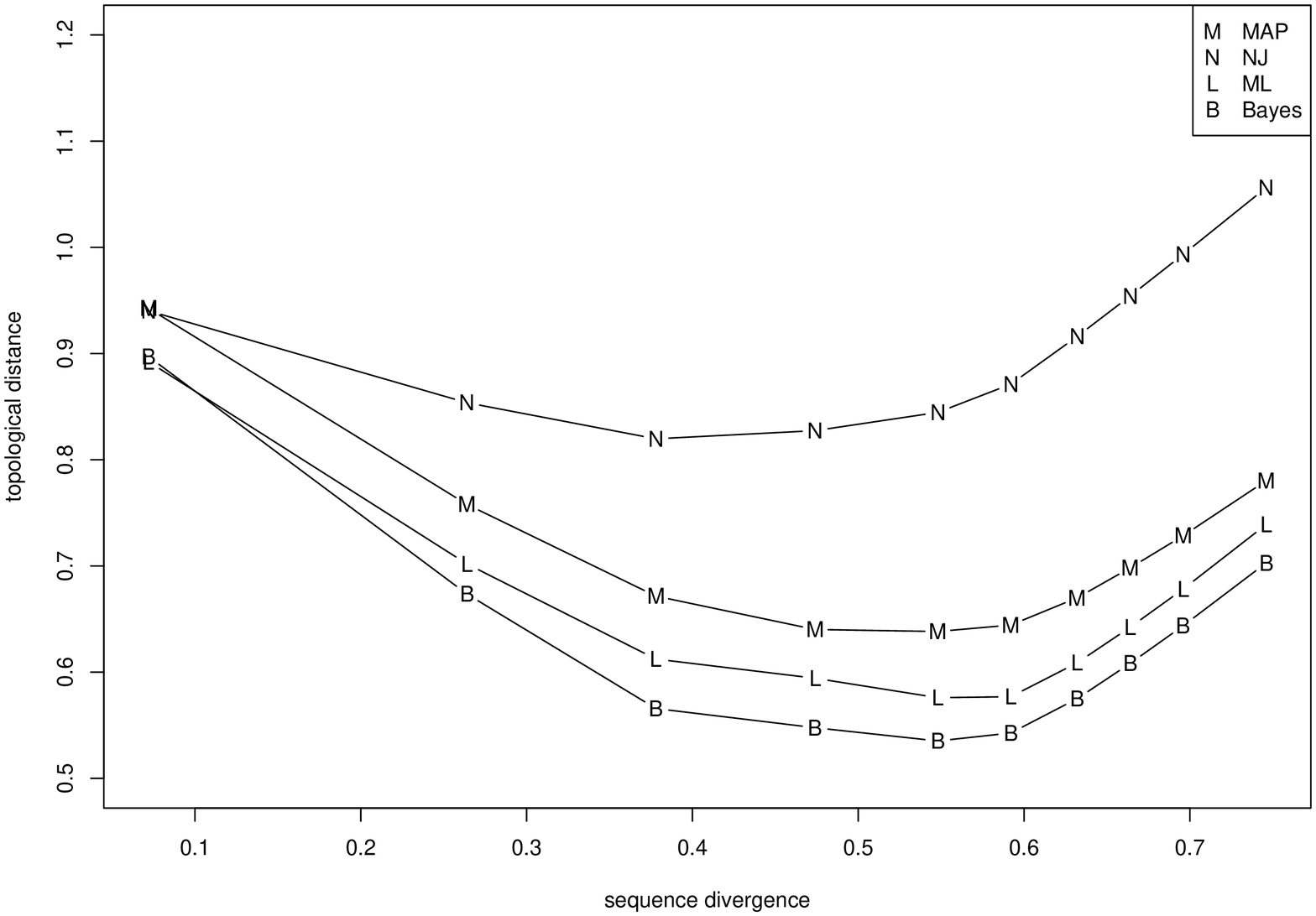}
\end{figure}

\pagebreak

{\bf Figure 2.}

\begin{figure}[ht!]
\includegraphics[scale= 0.6]{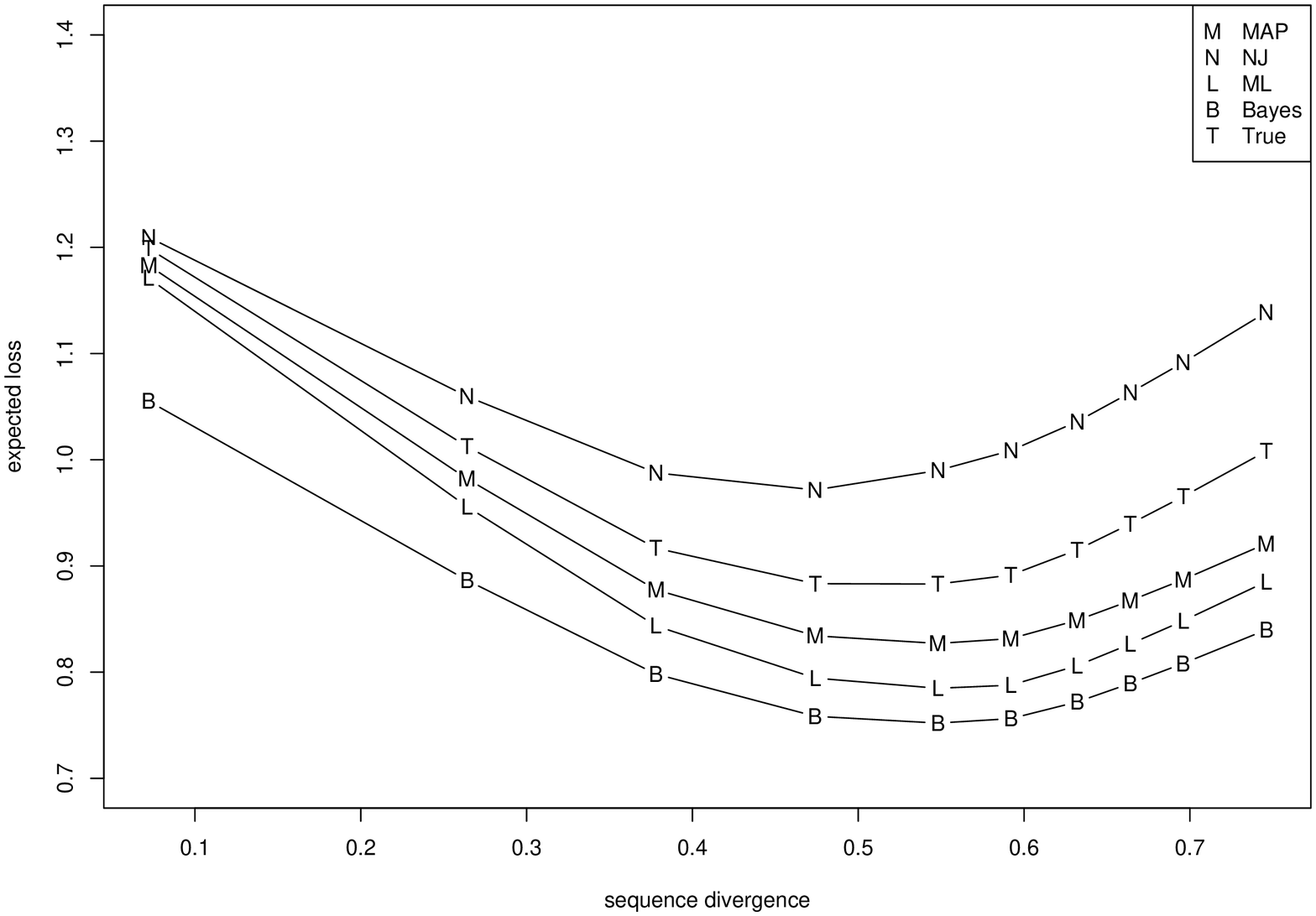}
\end{figure}

\pagebreak

{\bf Figure 3.}

\begin{figure}[ht!]
\includegraphics[scale=0.65]{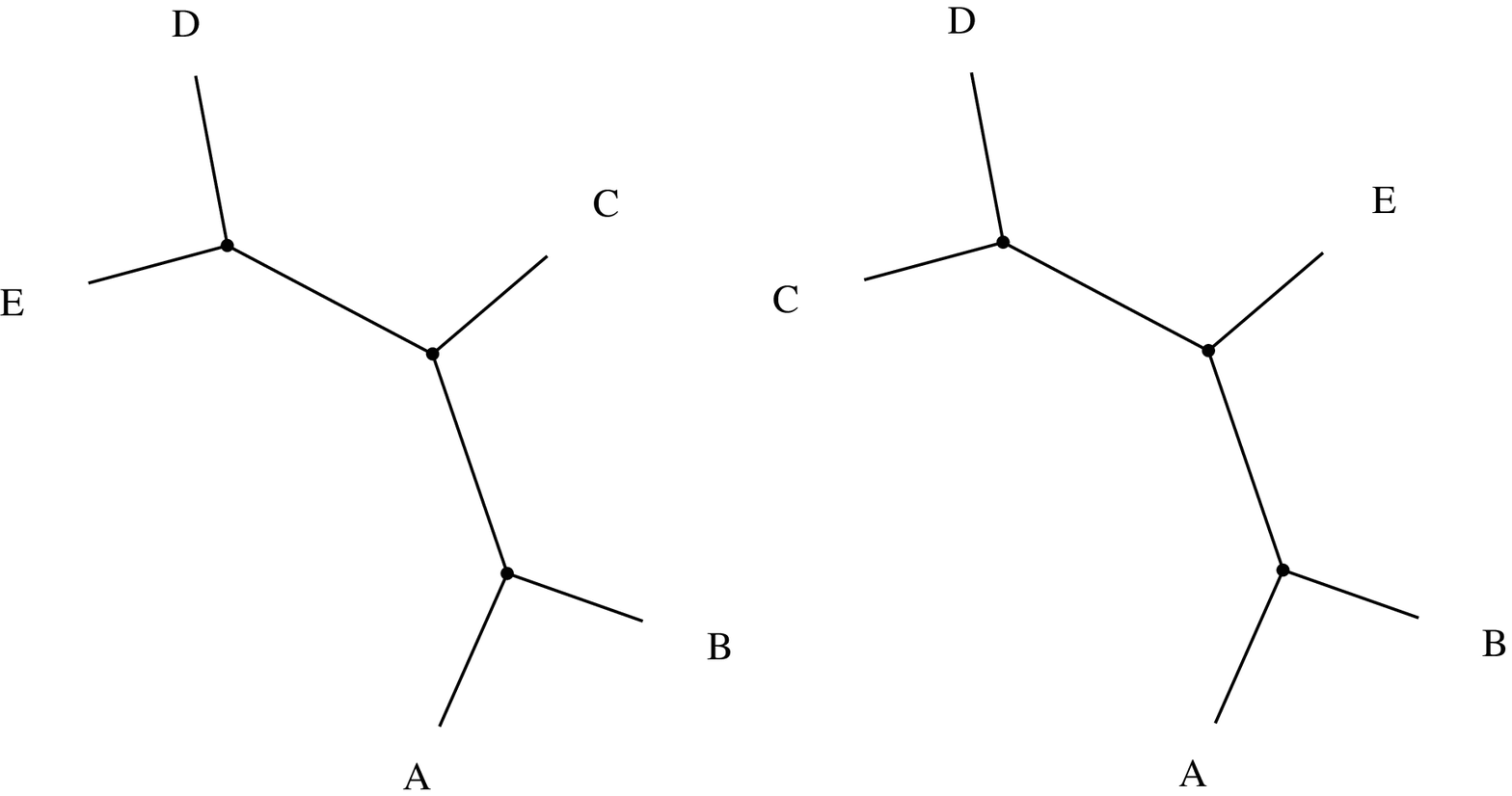}
\end{figure}

\pagebreak

{\bf Figure 4.}

\begin{figure}[ht!]
\includegraphics[scale=0.65]{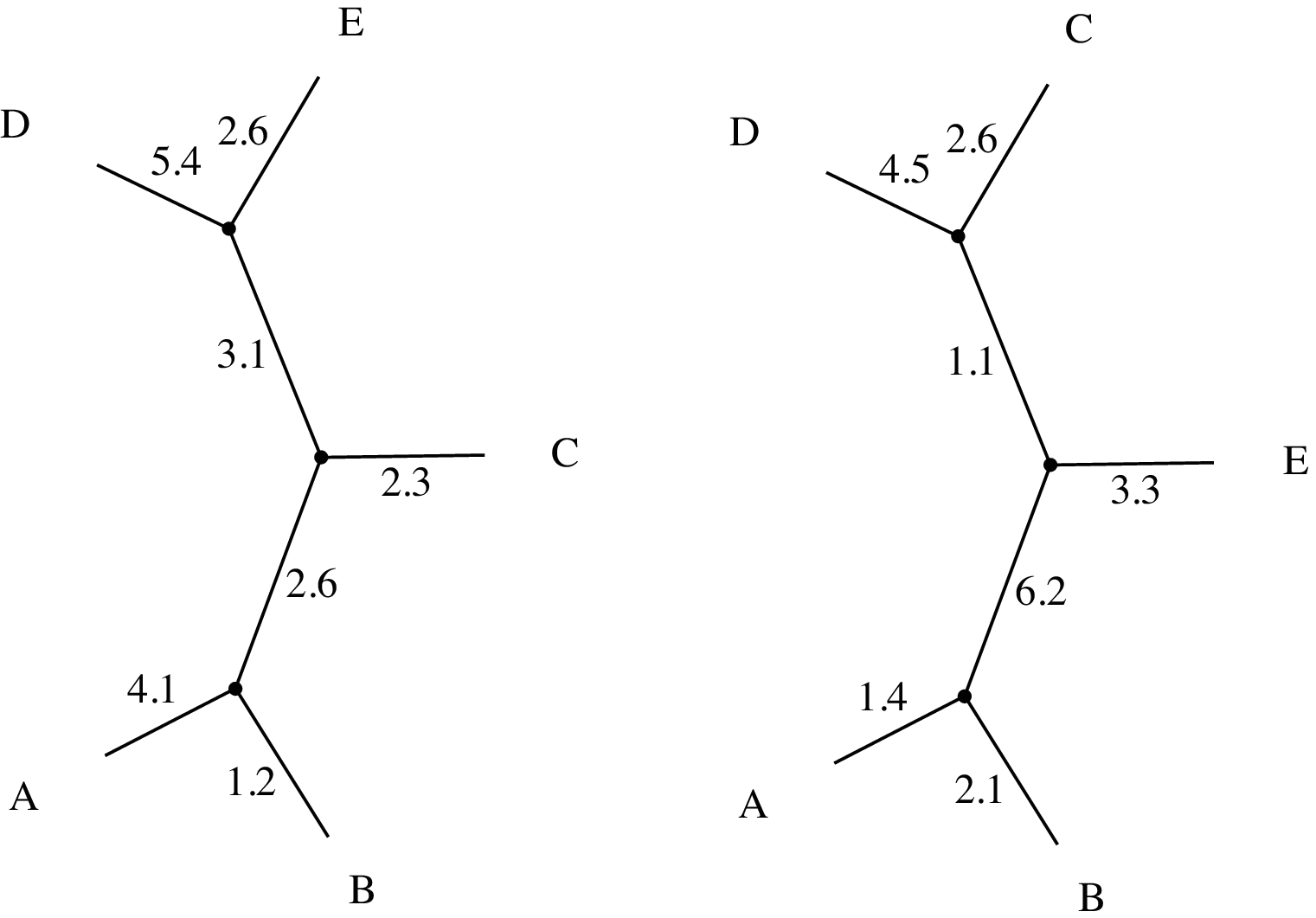}
\end{figure}

\end{document}